\documentclass[prc,preprint,superscriptaddress,showpacs,amssymb,amsmath,amsfonts,aps]{revtex4}

\usepackage{graphicx}% Include figure files
\usepackage{dcolumn}% Align table columns on decimal point
\usepackage{bm}% bold math
\usepackage{epsfig}% Include figure files

\begin{document}

\preprint{JLAB-PHY-07-744}

\title{Empirical Fit to Inelastic Electron-Deuteron and Electron-Neutron
Resonance Region Transverse Cross Sections}

\author{P.E.~Bosted}
     \email{bosted@jlab.org}
\affiliation{Jefferson Lab, Newport News, Virginia 23606}
\author{M.E.~Christy}
     \email{christy@jlab.org}
\affiliation{Hampton University, Hampton, Virginia 23668}

\date{\today}


\begin{abstract}
An empirical fit is described to  measurements of inclusive 
inelastic electron-deuteron cross sections in the kinematic range
of four-momentum transfer $0 \le Q^2<10$ GeV$^2$ and 
final state invariant mass $1.1<W<3.2$ GeV. The deuteron fit relies
on a fit of the ratio $R_p$ of longitudinal to transverse cross
sections for the proton, and the assumption
$R_p=R_n$. The underlying fit parameters describe the 
average cross section for a free proton and a free 
neutron, with a plane-wave
impulse approximation used to fit to the deuteron data. 
Additional fit parameters are used to fill in the dip between the
quasi-elastic peak and the $\Delta(1232)$ resonance. 
The mean deviation of data from the fit is 3\%, with less than
4\% of the data points deviating from the fit by more than 10\%. 
\end{abstract}
\maketitle

\section{Introduction}

Empirical knowledge of the inclusive electron-deuteron  cross 
section in the nucleon resonance region is important 
input for many research activities in nuclear and particle physics.  
The most important  examples include calculations 
of radiative corrections to cross sections, 
extractions of spin structure functions from asymmetry 
measurements, determinations of structure
function moments, and determinations of the vector 
coupling in models of low energy 
neutrino-nucleon cross sections.  The latter is of particular 
importance since the 
quality of low energy neutrino-nucleon cross section models will 
become one of the largest uncertainties in the 
extraction of neutrino oscillation parameters from 
future long-baseline experiments.

In this paper we will describe a fit to precision
electron-deuteron inelastic  cross sections in the resonance 
region for negative four-momentum transfer 
$0 \le Q^2$~$<$~10~${\rm GeV}^2$ and final state invariant 
mass $1.1<W<3.2$ GeV.  Among the advantages 
over previous fits~\cite{stuart,ioana_thesis} are:
inclusion of photoproduction and low $Q^2$ data points; inclusion
of several new experimental results; and the bulk of underlying fit is
to the free nucleon (average of free proton and neutron), with
Fermi motion consistently taken into account in a Plane Wave
Impulse Approximation (PWIA). The latter provides for a 
parameter-free way of describing the broadening of resonance
peaks with increasing three-momentum $\vec q$. The average nucleon
transverse cross section fit form is
the  same as the recent fit to electron-proton data of 
Ref.~\cite{Christy}, which uses a set of
threshold-dependent Breit-Wigner forms for all resonances.  
The principal difference is that the present fit is to the
transverse portion of the cross section only, due to a lack of
sufficient virtual photon polarization $\epsilon$ range in the
presently available deuteron data set. The assumption was
made that the ratio of longitudinal to transverse cross
sections, $R$, is the same for the free proton and neutron, as
supported by a recent analysis of data with $W>2$ GeV~~\cite{Tvaskis}, 
and also supported by the limited study of the present analysis. 
With the above assumptions, the
``dip'' region between the quasi-elastic peak and the $\Delta(1232)$
resonance is under-predicted at low $Q^2$, possibly because 
Meson Exchange Currents (MEC)
and Final State Interactions (FSI) have been ignored~\cite{laget}. 
An additional empirical function was used to fill in the missing
strength in the ``dip'' region.

The overall structure of the fit can be summarized as:
$$\sigma_D^T(W,Q^2) = \sigma_{dip}(W,Q^2) + \int 
 \sigma_N^T(W^\prime,(Q^2)^\prime) \Phi^2(\vec k) d^3 \vec k,$$
where $\sigma_D^T(W,Q^2)$ is the transverse cross section for
a deuteron,  $\sigma_{dip}(W,Q^2)$ is the dip region
parametrization (6 free parameters), and the integral is over
Fermi momentum $\vec k$ of the average free nucleon transverse cross
section $\sigma_N^T(W,Q^2)$ (36 free parameters). By simplifying
the 3-dimensional integration to a 1-dimensional integration
along the direction of the virtual photon, it was possible to
simultaneously fit all 42 parameters in a single gradient search
minimization. The minimization was done with respect to an
ensemble of approximately 15,000 data points from 9 experiments. 

The following section defines terms and kinematic variables. 
Section III describes the data sets used. Section IV gives
details of the functional form used for $\sigma_{dip}(W,Q^2)$
and $\sigma_N^T(W,Q^2)$, and how the Fermi-smearing integral
was simplified. The fit parameters are also listed in this
section. In Section V, we discuss various features of the results.

\section{Definitions and Kinematics}

In terms of the incident electron energy, $E$, the 
scattered electron energy, $E^{'}$, and the scattering angle, 
$\theta$, the absolute value of the exchanged 4-momentum squared 
in electron-nucleon scattering is given by
\begin{equation}
Q^2 = (-q)^2 =  4EE^{'}{\sin}^2 \frac{\theta}{2}, 
\end{equation}
and the mass of the undetected hadronic system is
\begin{equation}
W^2 = M_p^2 + 2M_p\nu -Q^2,  
\end{equation}
with $M_p$ the proton mass, $\nu = E-E^{\prime}$, and the small
terms involving the electron mass squared have been neglected. 

In the one-photon exchange approximation, the spin-independent 
cross section for inclusive electron-nucleon scattering can 
be expressed in terms of the photon helicity coupling as
\begin{equation}
\frac{d\sigma}{d\Omega dE^{'}} = \Gamma\left[
\sigma_N^T(W,Q^2) + \epsilon \sigma_N^L(W,Q^2)\right],
\label{eq:cs1}
\end{equation}
where $\sigma_N^T$ ($\sigma_N^L$) is the cross section for 
photo-absorption of purely
transverse (longitudinal) polarized photons,
\begin{equation}
\Gamma = \frac{\alpha E^{'}(W^2 - M_p^2)}{(2 \pi)^2 Q^2 M_p E (1 - \epsilon)}
\end{equation}
is the flux of transverse virtual photons, and
\begin{equation}
\epsilon = \left[1 + 2(1+\frac{\nu^2}{Q^2}) 
{\tan}^2 \frac{\theta}{2}\right]^{-1}
\end{equation}
is the relative flux of longitudinal virtual photons.
All the hadronic structure information is contained in 
$\sigma_N^T$ and $\sigma_N^L$, which are 
only dependent on $W$ and $Q^2$. We use the definition
$R_N=\sigma_N^L/\sigma_N^T$. We use the subscripts $p$, $n$,
$N$, and $D$ to refer to proton, neutron, average nucleon, and
deuteron respectively, where the deuteron cross sections are
defined to be per nucleon rather than per nucleus, 
following the high energy convention. We define 
$\sigma_N^{L,T}=(\sigma_p^{L,T}+\sigma_n^{L,T})/2.$

\section{Treatment of Experimental Data used in Fit}

\subsection{Description of the Data Sets}
The characteristics of the data sets used in the fit 
are summarized in Table~\ref{table_dataset}. The first
reference (Ref.~\cite{photo_armstrong}) includes
results from three early photoproduction 
experiments. We only 
used data from these experiments with 
beam energies above 800 MeV,
because at lower energies the more recent
photoproduction data of DAPHNE~\cite{daphne-p} has 
significantly smaller systematic errors. 
The JLab CLAS data~\cite{osipenko} cover a wide
kinematic range with many data points. The data are
reported as values of the structure function $F_2$,
averaged over two different beam energies. Since the
relative weight from the two beam energies (with 
differing values of $\epsilon$) was not given, there
is a systematic error in the conversion to $\sigma_T$
that was taken into account in the total error bars. 
The early JLab Hall C data of 
Niculescu~\cite{ioana_thesis} cover a similar kinematic
range as the CLAS data, with less data points but
smaller statistical and systematic errors. 
The more recent JLab Hall C  data of 
E00-116~\cite{simona1} cover the high $Q^2$ range
with relatively  few data points, which nonetheless have
very small statistical and systematic errors
(typically a few percent). To extend the low $W$
region to even higher $Q^2$, although with 
larger relative errors, we included the data
of SLAC E133~\cite{e133}. To cover the lower $Q^2$
region, we included preliminary data from two
JLab Hall C experiments: E02-109$^*$~\cite{Jan05} and 
E00-002~\cite{Edwin}. These experiments cover a wide
range of $\epsilon$ for each $(W,Q^2)$ point with
small statistical errors. Systematic errors were
being finalized at the time this fit was done, so
we used 2\% as a conservative estimate in lieu of the ultimate
errors for this experiment.

\begin{table}[tbh]
\begin{center}
\begin{tabular}{l l l l l c}
\hline
\hline
Data Set  & 
$Q^2_{Min}$  \hspace{0.5cm}  &  
$Q^2_{Max}$  \hspace{0.5cm}  &  
\# Data Points        \\
  &  ($\rm GeV^2$)  &  ($\rm GeV^2$) & \\  
\hline    
%
% both are set 1
Photoproduction (1972)~\cite{photo_armstrong} & 0 & 0 & 242 \\
Photoproduction (DAPHNE)~\cite{daphne-p} & 0 & 0 & 57  \\
%
% Clas set 2
CLAS~\cite{osipenko}                 &  0.35  &  5.9  & 11725  \\
%
% Ioana set 3
Early JLab~\cite{ioana_thesis}       &  0.50  &  4.2  &   600  \\
%
% Simona set 4
JLab E00-116~\cite{simona1}          &  3.6   &  7.5    & 288 \\
%
% E133 set 5
SLAC E133~\cite{e133}                &  2.5   &  10.0   & 488 \\
%
% Jan05 set 6
JLab E02-109$^*$~\cite{Jan05}             &  0.02   &  2.0    & 1435 \\
%
% Edwin set 7
 JLab E00-002$^*$~\cite{Edwin}            &  0.05   &   1.5    & 1445 \\
%
% E140 d2 cross sections
 SLAC E140~\cite{e140}                &  2.5    &  10.0    &   48 \\
\hline
\hline
\end{tabular}
\caption{Data sets used in fit. The number of data points and the
$Q^2$ range are indicated for each data set. $^*$The data from
Refs.~\protect{\cite{Jan05,Edwin}} are preliminary.}
\label{table_dataset}
\end{center}
\end{table}

In order to constrain the fit at high $W$ and $Q^2$, 
where there are insufficient data from JLab, we 
included the DIS data from SLAC E140~\cite{e140},  
and also added 
pseudo-data points from the SMC~\cite{SMC} fit to DIS $ed$ structure
functions. The pseudo-data were generated over the
interval $2.4<W<3.2$ GeV and $1.1<Q^2<10$ GeV$^2$. 
The errors used were those given by the SMC fit. 

The fit was found to be stable against the removal
of any particular data set.

\subsection{Quasi-elastic subtraction}
Inelastic electron scattering on the deuteron can be divided into two
distinct contributions: quasi-elastic scattering (just proton
and neutron in the final state), and inelastic scattering (one
or more mesons in the final state). Since the goal of the present
work is to fit inelastic scattering on the average free nucleon,
we have subtracted the 
quasi-elastic contribution (if not already done by the
experimenters) using the model described in Appendix I. 
Data points for which the quasi-elastic fraction was greater 
than 30\% of the total cross section were discarded,
as well as all points for which $W<1.1$ GeV.

\subsection{Longitudinal cross section  subtraction}
The next step in the data treatment  was to extract $\sigma^T_D$
from each of the electroproduction cross section measurements. 
The was done using:
$$\sigma^T_D = \sigma_D / (1 + \epsilon R_D).$$ 
As outlined in the introduction, we made the assumption that
$R_n = R_p$, and evaluated $R_D$ by Fermi-smearing both
$\sigma^L_p$ and $\sigma^T_p$ from the 
proton fit of Ref.~\cite{Christy}. The Fermi-smearing
procedure is described below. In practice, the 
Fermi-smearing had a very small effect for most
$(W,Q^2)$ points, so that,  to a good approximation,
$R_D=R_p$. 

\section{Description of the Fit}

\subsection{Overview}
A gradient-search minimization code (MINUIT) was used to 
simultaneously determine all 
fit parameters by minimizing 
the value of $\chi^2$, defined by:
$$\chi^2 = \sum_{i=1}^N [\sigma_i(W_i,Q^2_i) - 
\sigma_D^T(W,Q^2)]^2 / [\delta\sigma_i(W_i,Q^2_i)]^2$$
where the sum is overall experimental points 
with transverse
inelastic cross section $\sigma_i(W_i,Q^2_i)$
and total statistical and systematic error 
$\delta\sigma_i(W_i,Q^2_i)$. To avoid a tedious
iterative fit procedure, the Fermi-smearing 
integral in the model cross section 
$\sigma_D^T(W,Q^2)$ was simplified
to a single-dimension integral, as explained
below. Starting values of the average nucleon cross section
$\sigma^T_N(W^,Q^2)$ parameters were chosen to be the same as for the
equivalent parameters in the proton fit of Ref.~\cite{Christy}. 
Some adjustments of the upper and lower limits on the parameters
were needed to obtain the best $\chi^2$. The functional form
and starting parameters for the ``dip'' cross 
section $\sigma_{dip}(W,Q^2)$ were obtained from detailed study of
the fit residuals with the ``dip'' model absent.

In the next sections, we first describe the simplified Fermi-smearing
procedure, then the functional form of the free average nucleon
cross section, along with the fit parameters obtained, and finish
with the fit form and parameters for the ``dip'' region.

\subsection{Fermi-smearing}
The Fermi-motion of the nucleons in the deuteron was taken into
account using a PWIA calculation
and the Paris~\cite{Paris}  deuteron wave function
$\Phi^2(\vec k)$: 
\begin{equation}
\sigma_D(W,Q^2) = \int \sigma_N(W^\prime,(Q^2)^\prime) 
\Phi^2(\vec k) d^3 \vec k
\end{equation}
\noindent where we made the approximations:
\begin{equation}
(W^\prime)^2  = \left ( M_d  + \nu - \sqrt{M^2 + \vec k^2}\right )^2
- (\vec q)^2  - (\vec k)^2 + 2\vec q \cdot \vec k
\end{equation}
\noindent and $(Q^2)^\prime = Q^2$,
where $M_d$ is the deuteron mass, 
$M$ is the average nucleon  mass, and $\nu$ and 
$\vec q$ are the virtual photon energy and momentum, respectively.
Since these equations
basically boils down to the probability of finding a nucleon with
longitudinal  momentum
$k_z$, with the $z$ axis chosen along $\vec q$, we simplified
the problem by determining 20 values of $k^i_z$ for which the
integral over $\Phi^2(\vec k)$ is close to 1/20, in the
special case $\sigma_N(W^\prime,(Q^2)^\prime)=1$. The corresponding
average values of $(k^2)_i$ were also evaluated. The cross section
is then 
\begin{equation}
\sigma_D(W,Q^2) = \sum_{i=1}^{20} \sigma_N(W_i^\prime,(Q^2)^\prime)/20
\end{equation}
\noindent where now
\begin{equation}
(W_i^\prime)^2 = \left ( M_d  + \nu - \sqrt{M^2 + (k^2)_i} \right )^2
- (\vec q)^2  - (k^2)_i + 2 q k^i_z,
\end{equation}
This simplification is made possible by assuming that there
are no off-shell cross section corrections, and that the
$\vec k^2$ terms are small enough that the integral over
$\vec k^2$ at fixed $k_z$ can be replaced by evaluating
$W^\prime$ at the average values of $\vec k^2$. This approximation
works well for ranges in $W$ over which the cross section
is slowly varying. This is the case for the kinematic region
of the present fit, but is not advisable for smearing into
regions where the elementary cross section is zero (i.e. 
$W<M+M_\pi$). Hence, our fit is only valid for 
$W>1.1$ GeV. 

The numerical values of $k^i_z$ and $(k^2)_i$ 
are listed in Table~\ref{table_k}.

\begin{table}[tbh]
\begin{center}
\begin{tabular}{r r r r r r r r r r r r}
\hline
\hline
\hskip 0.1in $i=$ &\hskip 0.1in 1 & \hskip 0.1in 2 &  \hskip 0.1in 3 &  \hskip 0.1in 4 &  \hskip 0.1in 5 &  \hskip 0.1in 6 &  \hskip 0.1in 7 & \hskip 0.1in  8 & \hskip 0.1in  9 &  \hskip 0.1in 10 \\
\hline
$k_z^i$ & \hskip 0.1in  0.0029 & \hskip 0.1in 0.0083 & \hskip 0.1in 0.0139 & \hskip 0.1in 0.0199 & \hskip 0.1in0.0268 & \hskip 0.1in
 0.0349 & \hskip 0.1in 0.0453 & \hskip 0.1in 0.0598 & \hskip 0.1in 0.0844 & \hskip 0.1in 0.1853 \\
$(k^2)_i$ &
 0.0050 & 0.0051 & 0.0055 & 0.0060 &0.0069 &
  0.0081 & 0.0102 & 0.0140 & 0.0225 & 0.0964 \\
\hline
\hline
\end{tabular}
\end{center}
\caption{The first ten values of $k^i_z$ and $(k^2)_i$,
in units of GeV and GeV$^2$ respectively. 
The other 10 values are given by $k^{i+10}_z=-k^i_z$ 
and $(k^2)_{i+10} = (k^2)_i$.}
\label{table_k}
\end{table}

Due to the rapid variation of the cross section near threshold, 
we used 200 bins instead of 20 for $W<1.3$ GeV.

\subsection{Free Nucleon Fit Form}
The fit form used to describe $\sigma^T_N$ (the transverse
cross section for the average of a proton and a neutron) used
the same functional form as Ref.~\cite{Christy}.  
The total cross section is defined to be the incoherent sum of 
contributions from resonance production ($\sigma^R$) and a 
non-resonant background ($\sigma^{NR}$). 
The resonant cross section are described by threshold-dependent 
relativistic Breit-Wigner shapes with $Q^2$-dependent 
amplitudes for each resonance, such that
\begin{equation}
  \sigma^R (W^2,Q^2) = \sum_i BW^i (W^2) \cdot A_i^2(Q^2).  
\end{equation}
The form used for the Breit-Wigner resonance shapes  is given by 
\begin{equation}
%%%% BW^i = \frac {K_i K^{cm}_i}{ K K^{cm}} \cdot 
%%% added W
 BW^i = \frac {W K_i K^{cm}_i}{ K K^{cm}} \cdot 
\frac{\Gamma_i^{\rm tot} \Gamma_i^{\gamma} }{ \Gamma_i
\left [ (W^2 - M_i^2)^2 + (M_i \Gamma_i^{\rm tot})^2 \right ]}, 
\end{equation}
with 
\begin{equation}
K = {(W^2 - M_p^2) /  2 M_p},
\end{equation}
\begin{equation}
K^{cm} = {(W^2 - M_p^2) / 2 W}.
\end{equation}
Here, $K$ and $K^{cm}$ represent the equivalent photon 
energies in the lab and center of mass (CM) frames, respectively, 
while $K_i$ and $K^{cm}_i$ represent the same quantities 
evaluated at the mass of the $i^{th}$ resonance, 
$M_i$.  $\Gamma_i^{\rm tot}$ is the full decay width defined by 
\begin{equation}
\Gamma_i^{\rm tot} = \sum_j \beta^i_j \Gamma^i_j,
\end{equation}
with $\beta^i_j$ the branching fraction to the 
$\rm j^{th}$ decay mode for the $\rm i^{th}$ resonance 
and $\Gamma^i_j$ the partial width for this decay mode. 
The partial widths for single pion or eta decay 
were defined as
\begin{equation}
 \Gamma^i_j = \Gamma_i \left [\frac  { p^{cm}_j }{ p^{cm}_j |_{M_i}} 
 \right ]^{2L+1} \cdot \left [ \frac  { (p^{cm}_j)|_{M_i}^2 + 
 X_i^2}{ (p^{cm}_j)^2  + X_i^2} \right ]^L,
\end{equation}
where the $p^{cm}_j$ are meson momenta in the center of mass, 
$L$ is the angular momentum of the resonance, and 
$X_i$ is a damping parameter.  
For two-pion decays, we used:
\begin{equation}
 \Gamma^i_j = \frac {W \Gamma_i} {M_i} 
 \left [\frac  { p^{cm}_j }{ p^{cm}_j |_{M_i}} 
 \right ]^{2L+4} \cdot \left [ \frac  { (p^{cm}_j)|_{M_i}^2 + 
 X_i^2}{ (p^{cm}_j)^2  + X_i^2} \right ]^{L+2},
\end{equation}
% In the present analysis it was found that a good 
% fit to the data was obtained using $X_i = 0.1655$ for all the 
% resonaces except the $F_{15}(1680)$, where a larger value of 
% $X_i = 0.6$ was used. 
% below if only for proton, at present
%For the two-pion decay mode the 
%partial widths were determined from 
%\begin{equation}
%\Gamma^i_j = \Gamma_i \left [  \frac { p^{cm}_j }{ p^{cm}_j |_{M_i}} 
%\right ]^{2l+4} \cdot \left [ \frac { (p^{cm}_j)^2 + X_i^2}{ 
%(p^{cm}_j|_{M_i})^2  + X_i^2} \right ]^{l+2}.
%\end{equation}
The virtual-photon width was defined by:
\begin{equation}
\Gamma_i^\gamma = \Gamma_i  \left [ \frac { K^{cm} }{
 K^{cm} |_{M_i}} \right ]^2 \cdot 
\left [ \frac {(K^{cm}|_{M_i})^2  + X_i^2 }{
 (K^{cm})^2 + X_i^2} \right ]^2.
\end{equation}
Since $BW^i (W^2)$ depends only on $W^2$, it was
evaluated in 1 MeV bins in W from pion threshold to 5 GeV
and stored in a look-up table for future reference. This 
significantly reduced the time needed for $\chi^2$ evaluation
needed by the fitting code.

\begin{table}[tbh]
\begin{center}
\begin{tabular}{r r r r r r r r r r r r}
\hline
\hline
$i$ &  $M_i$ & $\Gamma_i$ & $L^i$ & $X_0^i$ & $\beta_{1\pi}^i$  
 & $\beta_{2\pi}^i$  & $\beta_{\eta}^i$ & 
%%% $A_i^2(0)$ & $c_1^i$ & $c_2^i$ & $c_3^i$ \\
%%% changed to be A
$A_i(0)$ & $c_1^i$ & $c_2^i$ & $c_3^i$ \\
\hline
\hskip 0.1in 
1 & \hskip 0.1in  1.230 & \hskip 0.1in  0.136 & \hskip 0.1in 1 & \hskip 0.1in 0.145 & \hskip 0.1in 1.00 & \hskip 0.1in 0.00 & \hskip 0.1in 0.00 & \hskip 0.1in  8.122 & \hskip 0.1in     5.19 & \hskip 0.1in      3.29 & \hskip 0.1in  1.870 \\
2 &  1.530 &  0.220 & 0 & 0.215 & 0.50 & 0.00 & 0.50 &  6.110 &   -34.64 &    900.00 &  1.717 \\
3 &  1.506 &  0.083 & 2 & 0.215 & 0.65 & 0.35 & 0.00 &  0.043 &   191.50 &      0.22 &  2.119 \\
4 &  1.698 &  0.096 & 3 & 0.215 & 0.65 & 0.35 & 0.00 &  2.088 &    -0.30 &      0.20 &  0.001 \\
5 &  1.665 &  0.109 & 0 & 0.215 & 0.40 & 0.60 & 0.00 &  0.023 &    -0.46 &      0.24 &  1.204 \\
6 &  1.433 &  0.379 & 1 & 0.215 & 0.65 & 0.35 & 0.00 &  0.023 &   541.90 &      0.22 &  2.168 \\
7 &  1.934 &  0.380 & 3 & 0.215 & 0.60 & 0.40 & 0.00 &  3.319 & 0 & 0 & 2.0 \\
\hline
\hline
\end{tabular}
\end{center}
\caption{Resonance parameters for states included in the fit. 
The branching ratios to single pion, double pion, and $\eta$ are denoted by 
$\beta_{1\pi}$, $\beta_{2\pi}$, and $\beta_{\eta}$ respectively. 
The assumed angular momentum is denoted by $L^i$. Units of
cross section are $\mu b$ and all masses, momenta, and energies
are in units of GeV.} 
\label{table_resparms}
\end{table}

For the transition amplitudes the fit form utilized was 
\begin{equation}
%%%  A_i^2(Q^2)  =  \frac { A_i^2(0)}{ (1 + Q^2/0.71)^{c_3^i} } \cdot 
% changed to 0.91, and not squared
  A_i(Q^2)  =  \frac { A_i(0)}{ (1 + Q^2/0.91)^{c_3^i} } \cdot 
\left( 1 + \frac {c_1^i Q^2 }{ (1 + c_2^i Q^2)} \right ). 
\end{equation} 
The parameters for all the resonances are listed 
in Table~\ref{table_resparms}. The variables $A_i(0)$, 
$c_1^i$, $c_2^i$, and $c_3^i$ were free parameters in the fit,
while all other parameters 
%% except the $\Delta(1232)$ mass and width
were fixed to those used in 
%% an early version of
the proton fit of Ref.~\cite{Christy}. 
% The final fit of 
% Ref.~\cite{Christy} differs from the earlier one: a) factor of 0.71
% changed to 0.91 in form for $A_i^2(Q^2)$; b) definition of
% $BW_i$ includes an extra factor of $W$; c) the partial width for
% two-pion decay has different powers involving angular momentum than the
% single-pion or $\eta$ decays. 
%Note that the present
%definition of $A_i^2(Q^2)$ is larger by a factor of $2M_p$ than
%the definition used in some other papers.

%\subsection{Non-resonant background}
The non-resonant background was parametrized as
\begin{equation}
\sigma^{NR} = \sum_{i=1}^2 x^\prime 
(C_1^i (\delta W)^{(2i+1)/2}) / (Q^2 + C_2^i)^{(C_3^i+C_4^i Q^2
+C_5^i Q^4)}
\end{equation}
where $\delta W = W - M_p - M_\pi$, $M_\pi$ is the pion mass,
and $x^\prime = 1 + (W^2 - (M_p + M_\pi)^2)/(Q^2 + C_6)$, and
the fit parameter $C_6=0.05$ GeV$^2$. The fit values for
the other ten parameters are listed in Table~\ref{table_nrparms}.

\begin{table}[tbh]
\begin{center}
\begin{tabular}{r r r r r r}
\hline
\hline
i & $C^i_1$ & $C^i_2$ & $C^i_3$ & $C^i_4$ & $C^i_5$ \\
\hline
\hskip 0.1in 
1 &\hskip 0.1in  226.6 & \hskip 0.1in 0.0764 & \hskip 0.1in 1.4570 & \hskip 0.1in  0.1318 & \hskip 0.1in -0.005596 \\
2 & -75.3 &  0.1776 &  1.6360 &  0.1350 &  0.005883 \\
\hline
\hline
\end{tabular}
\end{center}
\caption{Non-resonant parameters as described in the text.
Units of cross section are $\mu b$ and all masses, momenta, 
and energies are in units of GeV.}
\label{table_nrparms}
\end{table}

\subsection{Dip region parametrization}
It was found that with the assumption that the PWIA ``smearing'' of
free nucleon cross sections describes electron-deuteron scattering,
there was always missing strength in the ``dip'' region between
the quasi-elastic peak at $W=M_p$, and the low $W$ side of the
$\Delta(1232)$ resonance. This
missing strength could be due to MEC and FSI in either quasi-elastic
or inelastic scattering. Whatever the physical cause, we chose
a purely empirical form that greatly improved the fit quality.
This dip region additive correction term is given by: 
\begin{equation}
F_1^{\rm dip}=1.964
\nu^{0.298} e^{[-(W-1.086)^2/0.00531]}/(1 + {\rm max}(0.3,Q^2)/1.265)^8
\end{equation}
The unit-less structure function $F_1$ is related to 
$\sigma_T$ by
\begin{equation}
F_1= (W^2-M_p^2) \sigma_T / 8 \pi^2 \alpha (\hbar  c)^2
\label{Eq:sigtof1}
\end{equation}
\noindent where $\alpha$ is the fine structure constant.

\section{Fit results}
\begin{figure}[hbt]
\centerline{\epsfig{file=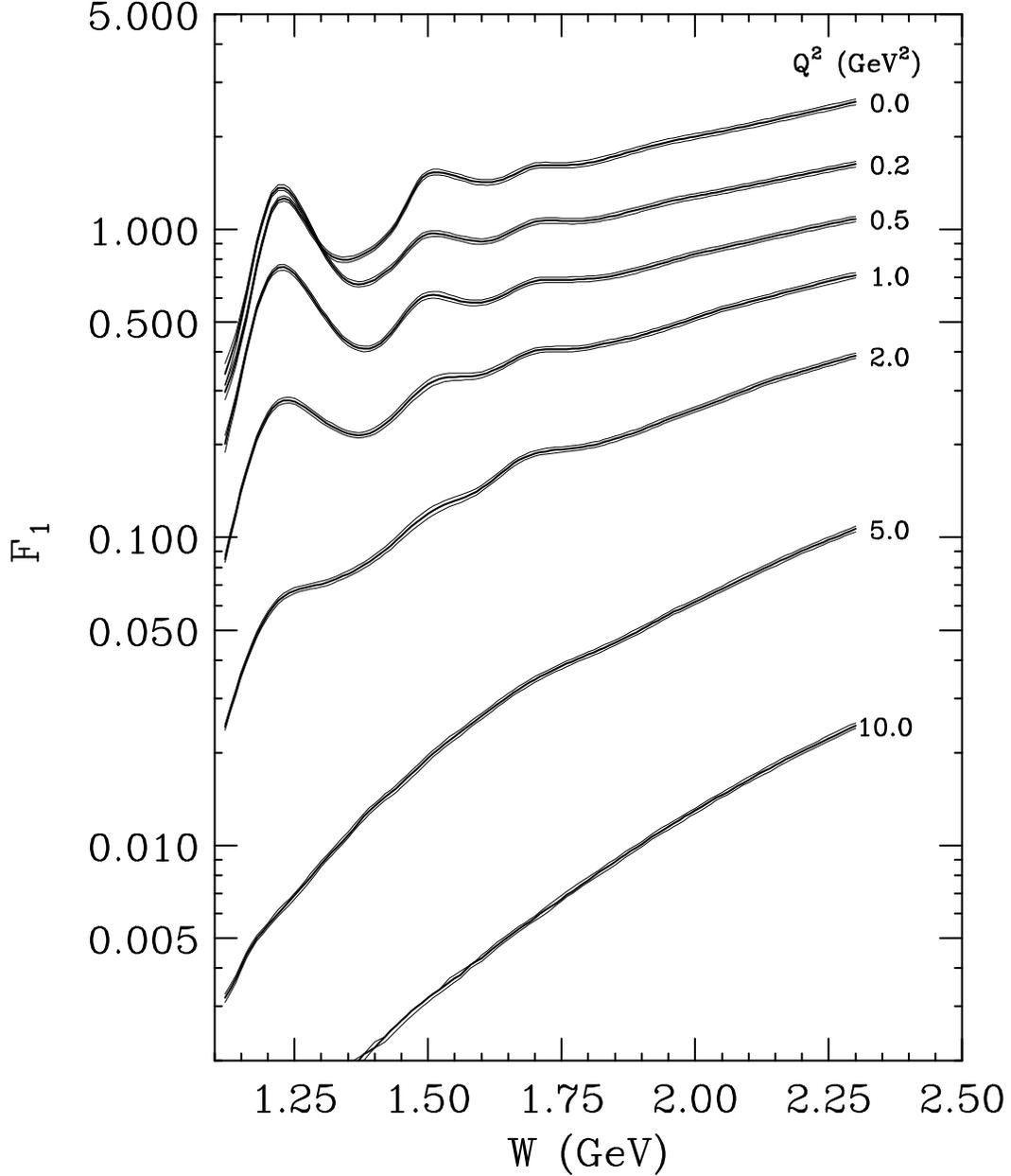,width=5.5in}}
\caption{Fit results for $F_1$/nucleon for the deuteron
as a function of $W$ at representative values of 
$Q^2$ as indicated on the figure. Central values are shown as
the thick lines, while the 1-$\sigma$ error bands are shown 
as the thin lines.}
\label{fig:fig1}
\end{figure}

The results of the deuteron fit are shown at seven representative
values of $Q^2$ as a function of $W$ in Fig.~\ref{fig:fig1}, along
with the 1-$\sigma$ error band. 
In order to span a smaller range on the vertical axis, we
plot $F_1$ rather than $\sigma_T$.
% Note the strong dependence on $W$ and $Q^2$ in the threshold region
% which makes a continuous fit problematic. Also n
Note that the
resonant structure, clearly visible at low $Q^2$, is essentially
gone for $Q^2>5$ GeV$^2$, due to the increasing influence of
Fermi broadening and non-resonant background contributions.

\begin{figure}[hbt]
\centerline{\epsfig{file=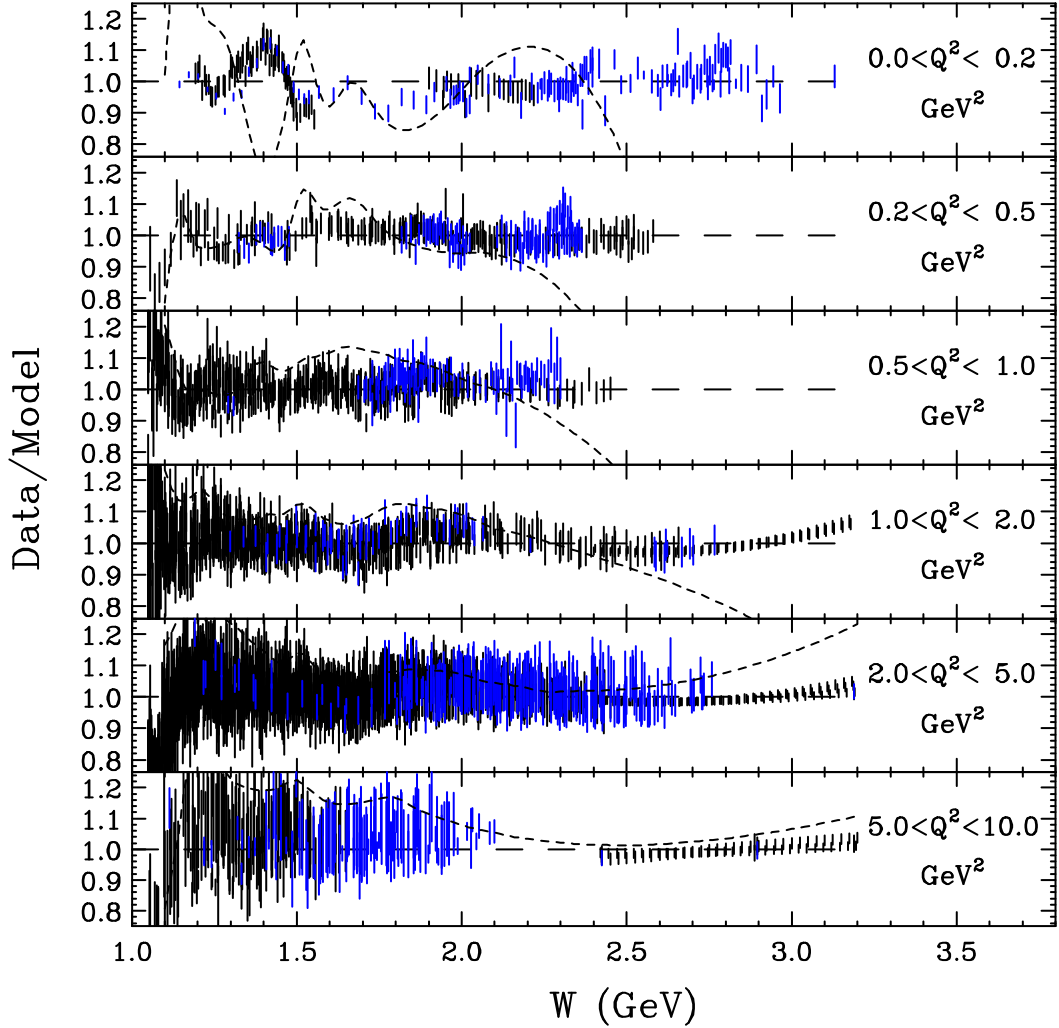,width=5.5in}}
\caption{Ratio of $ed$ inelastic 
data to model as a function of $W$ in
six ranges of $Q^2$. The gray (blue online) points correspond
to $\epsilon<0.5$, while the black points are for $\epsilon>0.5$.
% The points for $W<1.2$ GeV are pseudo-data for $eN$ scattering
% from MAID 2007~\protect{\cite{MAID}}. 
Most of the points
for $W>2.4$ GeV and $Q^2>1$ GeV$^2$ are from the $ed$ fit from
SMC~\protect{\cite{SMC}}. The ratio of the fit of 
Niculescu~\protect{\cite{ioana_thesis}} to the present fit is illustrated
by the dashed curves. }
\label{fig:fig2}
\end{figure}

\begin{figure}[hbt]
\centerline{\epsfig{file=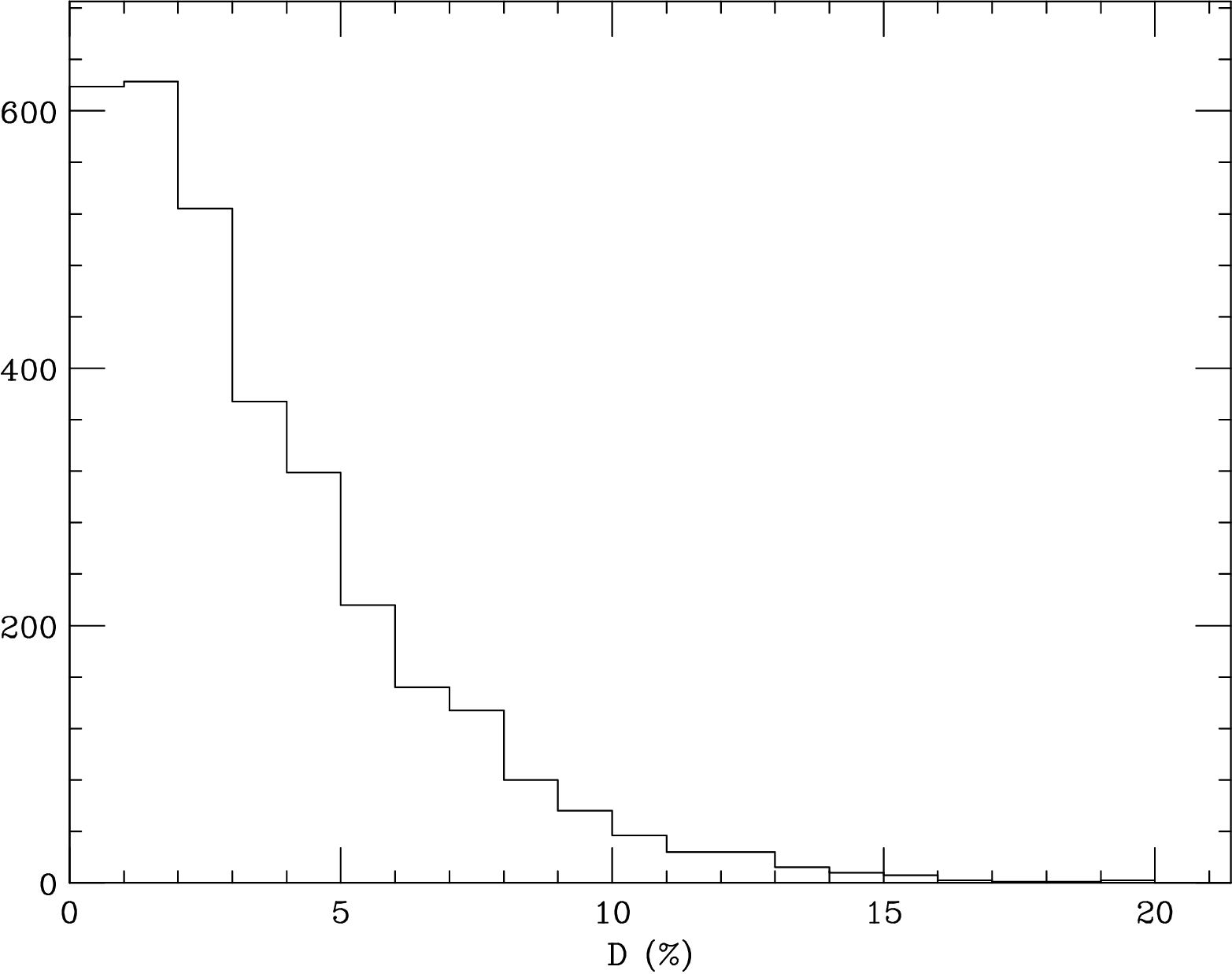,width=5.5in}}
\caption{Frequency distribution for the deviations from unity of
the ratios of data to fit.}
\label{fig:fig3}
\end{figure}

Ratios of cross sections for each data point to the corresponding
fit value are shown in Fig.~\ref{fig:fig2} as a function of 
$W$ for six bins in $Q^2$. As shown in  Fig.~\ref{fig:fig3}, 
96\% of the points lie within 10\% of the fit, 76\% lie within
5\%, and 53\% lie within 3\%. 
% Only data points with statisitcal
% errors less than 5\% were used in this plot. 
Overall, the
agreement of data and fit is reasonably good at the 3\% to 5\%
level. It can be seen in Fig.~\ref{fig:fig2} 
that there are two noticeable oscillations in
the ratios at low $W$ and $Q^2$: this trend is
also seen in the proton fit of Ref.~\cite{Christy}. 
This may indicate the need for additional parameters to describe
the photoproduction and very low $Q^2$ data.
%There is also an oscillation in the ratios near
%$W=1.2$ GeV for $Q^2>1$ GeV$^2$. This may be due to the assumption in our
%fit of no interference terms between the $\Delta(1232)$ resonance
%and the non-resonant background. 

In order to check the assumption that $R_p=R_n$, the data
with $\epsilon<0.5$ are plotted in gray (blue online), while
those with $\epsilon>0.5$ are plotted in black. No glaring
discrepancy is seen between these two data sets. A more
refined analysis will be performed once the preliminary
data of JLab  E02-109~\cite{Jan05} and 
E00-002~\cite{Edwin}  are finalized, 
and results from the very recent JLab  
E06-009~\cite{Rosen07} experiment become available. 
The E02-109 and E06-009 experiments were 
specifically designed to
measure $R_d$, and cover a large range of $\epsilon$ at 
many specific values of $(W,Q^2)$. 

The ratio of the commonly-used fit of Niculescu~\cite{ioana_thesis}
to the present fit are also shown in Fig.~\ref{fig:fig2}. 
That fit tends to systematically lie above the data in the
resonance region ($W<2$ GeV), and lies well below data
at high $W$ and low $Q^2$ (outside the kinematic 
range of that fit). 

\begin{figure}[hbt]
\centerline{\epsfig{file=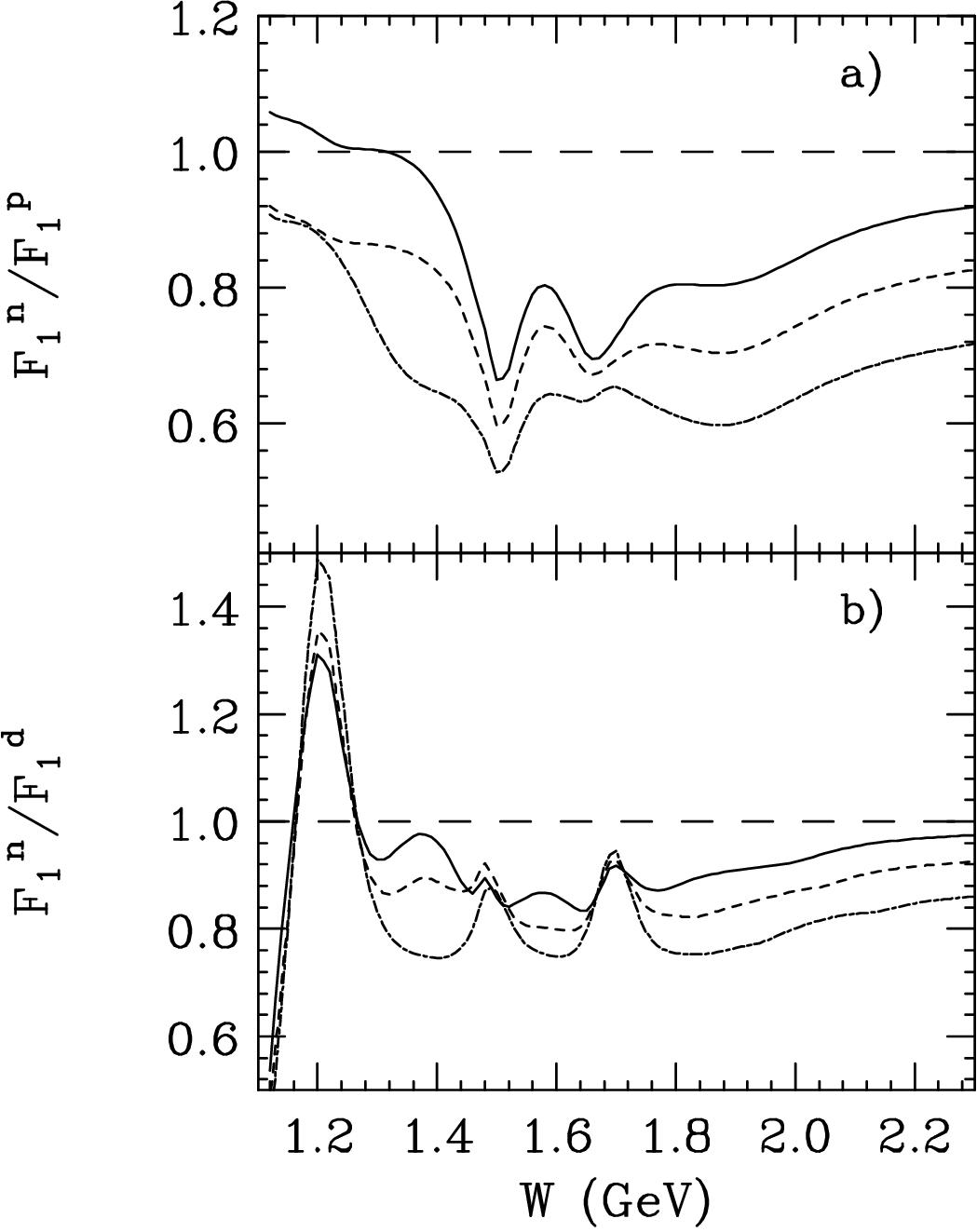,width=5.5in}}
\caption{Fit results for a) the ratio $F_1^n/F_1^p$ 
and b) the ratio $F_1^n/F_1^d$ as a 
function of $W$ for $Q^2=0.5$ GeV$^2$ (solid curves), 
$Q^2=1$ GeV$^2$ (dashed curves), and 
$Q^2=2$ GeV$^2$ (dot-dashed curves). $F_1^d$ is defined to be
per nucleon.} 
\label{fig:fig4}
\end{figure}

Since the present underlying fit is to the average free nucleon, we can use
the proton fit~\cite{Christy} to obtain predictions for the
ratio of neutron to proton transverse cross sections (or
equivalently the ratio $F_1^n/F_1^p$), as illustrated in
Fig.\ref{fig:fig4}a. Significant  resonance structure
is predicted, especially at low $Q^2$ and in the region of 
the $\Delta(1232)$ resonance,
for which the resonant contribution to $F_1^n/F_1^p$ is 
expected to be unity 
by isospin invariance. 
%There is also a large oscillation predicted
%near $W=1.4$ GeV, which may be due to a strong isospin
%dependence to the Roper resonance.
These predictions can be
tested against the anticipated results of 
the JLab ``BONUS'' experiment~\cite{Bonus},
which used tagging of low energy backward protons to ``tag''
spectator protons in electron-deuteron scattering (and hence
isolate electron-neutron scattering). Predicted ratios
of $F_1^n/F_1^d$ that could be extracted from BONUS are shown
in Fig.\ref{fig:fig4}b at three representative values of $Q^2$.

\section{Summary}
An empirical fit to inelastic electron-deuteron scattering has been
performed which describes available data reasonably well 
(3\% to 5\% level) in nearly all of 
the kinematic range 
than can be accessed  at
Jefferson Lab with up to 6 GeV electrons and photons: 
$0 \le Q^2<10$ GeV$^2$ and $1.1<W<3.2$ GeV. 
%The fit can also be used
%between single and double pion threshold with an overall accuracy
%that decreases from about 5\% at low $Q^2$ to about 15\% at
%$Q^2=5$ GeV$^2$. 
The fit is useful in the evaluation of radiative
corrections to experimental data, for extraction of spin structure
functions from asymmetry measurements, and for the evaluation of
structure function moments. Since the underlying fit is to an
average nucleon, the results can be combined with a proton 
fit~\cite{Christy} to obtain predictions for electron-neutron
scattering in the resonance region. Suitably corrected for Fermi
motion, these can in turn be used to make neutron excess  corrections
to nuclear structure functions. 

Once the data from JLab  E02-109~\cite{Jan05}, E00-002~\cite{Edwin},
and JLab  E06-009~\cite{Rosen07} are finalized, we plan to
re-do the fit for $\sigma_L$ and $\sigma_T$ separately, rather
than making the assumption $R_p=R_n$. We also plan to use the
results of the BONUS~\cite{Bonus} experiment, which to first order
will measure the ratio of electron-neutron to electron-deuteron
scattering. 

FORTRAN computer code embodying the electron-deuteron fit described in
this article is available by email request from the
authors. The code includes the fit covariance matrix (or
error matrix), and a subroutine to obtain the error on
$F_1^d$. The code also includes the proton
fit of Ref.~\cite{Christy}, permitting the determination of electron-neutron
cross sections from the deuteron and proton fits. 
The same code also includes
a simple Fermi-smearing model of electron-nucleus cross sections
for $A>2$, using these deuteron and proton
fits, as described in Ref.~\cite{cn15}.
The quasi-elastic model described in the Appendix is also included.

%Finally, a self-consistent set of radiative corrections should improve
%the agreement among experiments in the region near pion threshold,
%and, after accounting for experimental resolution, allow for an
%empirical fit to better describe the effects of MEC, FSI, and
%interference terms that tend to fill in the dip region between
%quasi-elastic scattering and the $\Delta(1232)$ resonance. 

\begin{acknowledgments}
This work was supported in part by research grants 0099540 and 9633750 from the National Science Foundation. The Southeastern
Universities Research Association (SURA) operated the Thomas
Jefferson National Accelerator Facility for the United States
Department of Energy under contract DE-AC05-84ER40150. 
\end{acknowledgments}

\section{Appendix I: Quasi-elastic Model}
\begin{figure}[hbt]
\centerline{\epsfig{file=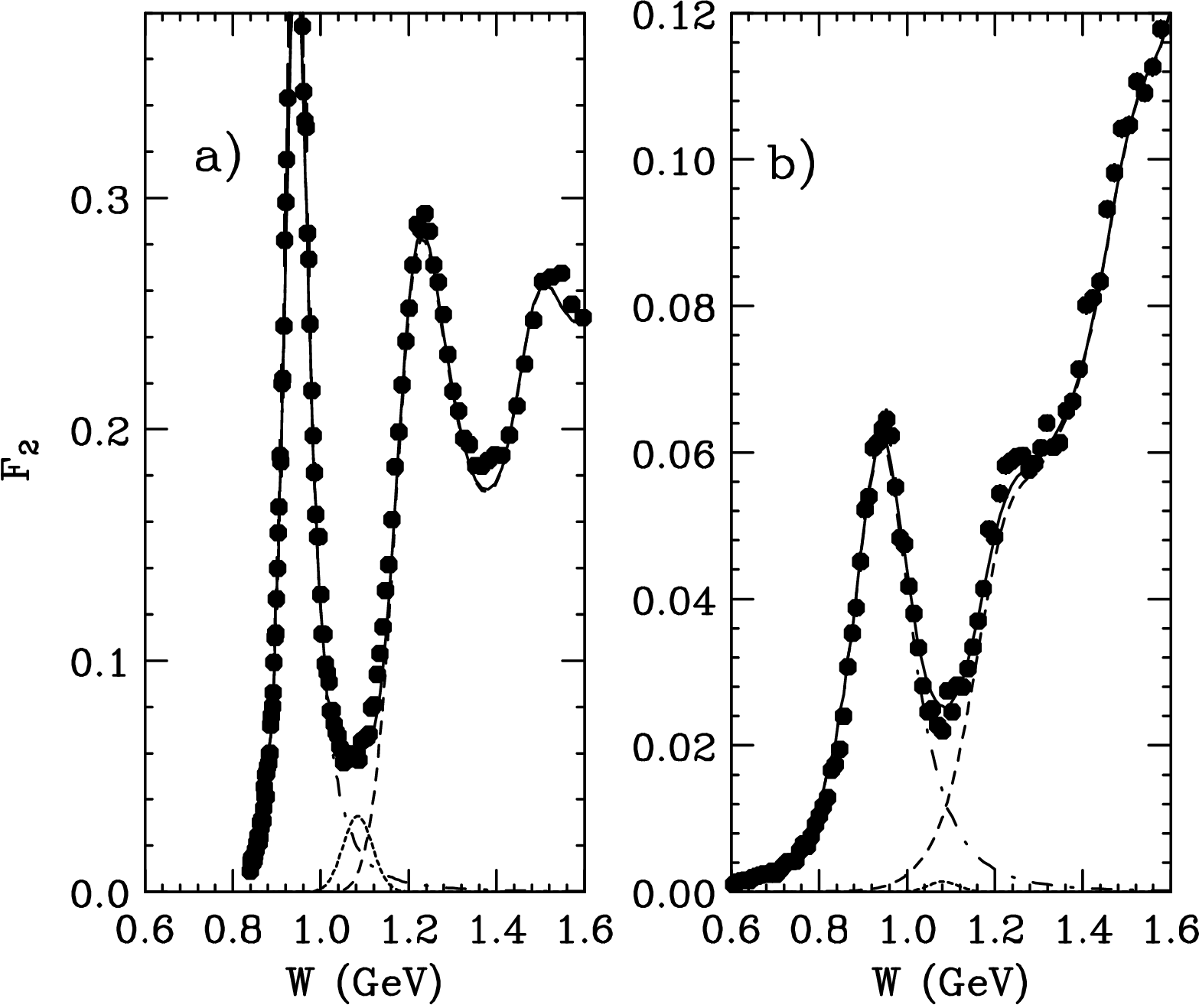,width=5.5in}}
\caption{Comparison of the quasi-elastic model (dot-dashed
curves), PWIA part of the inelastic model (dashed curves),
``dip region'' part of the inelastic model (short dashed curves), 
 and their sum (solid
curves) with $F_2$ data from Ref.~\protect{\cite{osipenko}} at 
a) $Q^2=0.525$ 
GeV$^2$ and b) $Q^2=2.075$ GeV$^2$.}
\label{fig:qe}
\end{figure}

To model quasi-elastic scattering, we used the same
PWIA Fermi-smearing prescription (based on the
deuteron Paris wave function) as for inelastic
scattering, except that the continuous inelastic
cross section was replaced by a $\delta$-function
elastic cross at $W=M_p$. The elastic cross section
was calculated using the nucleon form factors
of Bosted~\cite{Bosted}, modified for off-shell
effects using the prescription of Ref.~\cite{Donnelly}.
Following Tsai~\cite{tsai}, Pauli suppression was
taken into account using the factor
$(3 q / 4k_f){1 - [(q / k_f)^2]/12}$, for $q<2k_f$, 
and unity for $q>2k_f$, where $q$ is the magnitude
of $\vec q$, and we used $k_f=0.085$ GeV. As shown in
Fig.~\ref{fig:qe}, the model works reasonably well
near the quasi-elastic peak at two values of $Q^2$. 
It can also be seen that the empirical ``dip region'' parametrization
is useful to describe the region near $W=1.09$ GeV at low $Q^2$.

%\bibliography{2002}

\clearpage

\end{document}